\documentclass[preprint,showpacs,preprintnumbers,amsmath,amssymb]{revtex4-1}

\usepackage{graphicx}
\usepackage{amssymb}

\newcommand{\Q}{\mathbf{Q}}

\newcommand{\PP}{\mathbf{P}}
\newcommand{\rr}{\mathbf{r}}

\newcommand{\kk}{\mathbf{k}}

\newcommand{\e}{\mathbf{e}}
\newcommand{\z}{\mathbf{\hat z}}
\newcommand{\nn}{\mathbf{\hat n}}
\newcommand{\mm}{\mathbf{\hat m}}

\begin{document}

\title{Nematic twist--bend phase in an external field}

% Use letters for affiliations, numbers to show equal authorship (if applicable) and to indicate the corresponding author

\author{Grzegorz Paj\c{a}k}
\email[e-mail address:]{grzegorz@th.if.uj.edu.pl}
\affiliation{Marian Smoluchowski Institute of Physics, Department
of Statistical Physics, Jagiellonian University, \L{}ojasiewicza 11,
Krak\'ow, Poland \\
and \\
Faculty of Physics, Mathematics and Computer Science,
Tadeusz Ko\'sciuszko Cracow University of Technology, Podchor\c{a}\.{z}ych 1, 30-084 Krak\'{o}w, Poland }
\author{Lech Longa}
\email[e-mail address:]{lech.longa@uj.edu.pl}
\affiliation{Marian Smoluchowski Institute of Physics, Department
of Statistical Physics, Jagiellonian University, \L{}ojasiewicza 11,
Krak\'ow, Poland}
\author{Agnieszka Chrzanowska}
%\email[e-mail address:]{lech.longa@uj.edu.pl}
\affiliation{Faculty of Physics, Mathematics and Computer Science,
Tadeusz Ko\'sciuszko Cracow University of Technology, Podchor\c{a}\.{z}ych 1, 30-084 Krak\'{o}w, Poland}

 \date{\today}

\begin{abstract}
The response of the nematic twist--bend ($N_{\textrm{TB}}$) phase to an applied field can provide important insight into structure of this liquid and  may bring us closer to understanding  mechanisms generating mirror symmetry breaking in a fluid of achiral molecules. Here we investigate theoretically how an external uniform field can affect
structural properties and stability of $N_{\textrm{TB}}$. Assuming that the driving force responsible for the formation of this phase is packing entropy we show, within Landau--de Gennes theory, that $N_{\textrm{TB}}$ can undergo a rich sequence of structural changes with field. For the systems with positive anisotropy of permittivity we first observe a decrease of the tilt angle of  $N_{\textrm{TB}}$ until it transforms through a field--induced phase transition to the ordinary prolate uniaxial nematic phase ($N$). Then, at very high  fields this nematic phase develops  polarization perpendicular to the field.
For systems with negative anisotropy of permittivity the results reveal new modulated structures. Even an infinitesimally small field transforms $N_{\textrm{TB}}$  to its elliptical counterpart ($N_{\textrm{TB} e}$), where the circular base of the cone of the main director becomes elliptic. With stronger fields the ellipse degenerates to a line giving rise to a nonchiral periodic structure, the nematic splay-bend ($N_{\textrm{SB}}$), where the two nematic directors are restricted to a plane. The three structures, $N_{\textrm{TB}}$, $N_{\textrm{TB} e}$ and $N_{\textrm{SB}}$, with a modulated polar order are globally nonpolar. But further increase of the field induces phase transitions into globally polar structures with nonvanishing polarization along the field's direction. We found two such structures, one of which is a polar and chiral modification of $N_{\textrm{SB}}$, where splay and bend deformations are accompanied by weak twist deformations ($N^*_{\textrm{SB} p}$). Further increase of the field unwinds this structure into a polar nematic ($N_p$) of polarization parallel to the field.
\end{abstract}
\maketitle

%\dates{This manuscript was compiled on \today}
%\doi{\url{www.pnas.org/cgi/doi/10.1073/pnas.XXXXXXXXXX}}

% Optional adjustment to line up main text (after abstract) of first page with line numbers, when using both lineno and twocolumn options.
% You should only change this length when you've finalised the article contents.
%\verticaladjustment{-2pt}

%\maketitle
%\thispagestyle{firststyle}
%\ifthenelse{\boolean{shortarticle}}{\ifthenelse{\boolean{singlecolumn}}{\abscontentformatted}{\abscontent}}{}

% If your first paragraph (i.e. with the \dropcap) contains a list environment (quote, quotation, theorem, definition, enumerate, itemize...), the line after the list may have some extra indentation. If this is the case, add \parshape=0 to the end of the list environment.
%\dropcap
{T}he twist--bend nematic ($N_{\textrm{TB}}$) phase, recently discovered in liquid--crystalline chemically achiral dimers \cite{Sepelj&Lesac2007,Panov&Nagaraj2010,Cestari&DiezBerart2011,Borshch&Kim2013,Chen&Porada2013,%
Paterson&Gao2016,Lopez&RoblesHernandez2016}, bent--core mesogens \cite{Gortz&Southern2009,Chen&Nakata2014}, and their hybrids \cite{Wang&Singh2015}, is one of the most amazing example of spontaneous chiral symmetry breaking in soft matter physics. It occurs in the liquid state without any long--range positional order, but the average local molecular long axis, $\mathbf{\hat{n}}$, known as the director, follows a nanoscale--pitch heliconical winding. Thus the structure belongs to the family of nematic phases and is the fifth nematic phase recognized \cite{Chen&Porada2013}, in addition to uniaxial and biaxial nematics for nonchiral liquid crystalline materials and cholesteric and blue phases for chiral liquid crystals \cite{deGennesBook}. In 2001 Dozov \cite{Dozov2001}, following earlier analysis of Meyer \cite{Meyer1969,MeyerBook}, predicted theoretically this structure using Frank model of elastic deformations in nematics by assuming that the bend elastic constant can change sign. With this assumption $\mathbf{\hat{n}}$ can form 1D modulated structures where simultaneously twist and bend or splay and bend elastic constants are nonzero. The second of the structures, known as the nematic splay--bend ($N_{\textrm{SB}}$), is nonchiral and exhibits periodic splay and bend modulations of the director, taking place within one plane. The observation of this phase is still not confirmed experimentally, but it can be stabilized in constant pressure Monte--Carlo simulations of thin layers composed of hard bent--core molecules \cite{Karbowniczek&lc&2017}.

\begin{figure}[h!]
\centering
\includegraphics[scale=0.51]{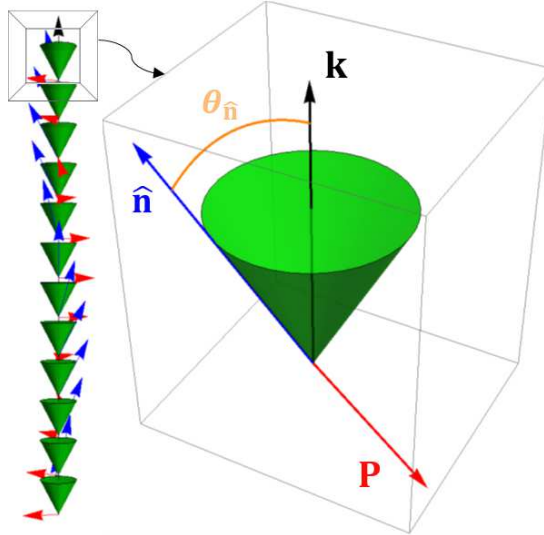}
\caption{(Color online) Schematic representation of the nematic twist--bend structure. Right circular cone of angle $\theta_\nn$ shows
the tilt between the director $\nn$ and the helical symmetry axis, parallel to the wave vector, $\kk$.
Red arrows represent $\PP$, where $\PP \parallel \kk \times \nn $;  black arrow is the direction of $\kk$. }
\label{Fig1}
\end{figure}

The first possibility is recognized as the chiral $N_{\textrm{TB}}$ phase with the director, $\mathbf{\hat{n}}(\mathbf{r}) \equiv \mathbf{\hat{n}}(z)$, attaining oblique helicoidal structure in precessing on the side of a right circular cone, Fig.~\ref{Fig1}.
More specifically
\begin{equation}\label{nn}
 \mathbf{\hat{n}}(z) = [\cos (\phi_\nn) \sin (\theta_\nn), \sin (\phi_\nn) \sin (\theta_\nn), \cos (\theta_\nn)],
\end{equation}
where $\theta_\nn$ is the conical angle (angle between $\mathbf{\hat{n}}$ and $\mathbf{k}$) and $\phi_\nn= \pm k z = \pm \frac{2\pi}{p} z$ with wave vector $\mathbf{k}=k \mathbf{\hat{z}}$ taken to be parallel to the $\mathbf{\hat{z}}$--axis of laboratory system of frame; here $p$ is the pitch. The Dozov's scenario for the formation of $N_{\textrm{TB}}$ has strong experimental support for anomalously small, but positive values of the bend elastic constant have been reported in the nematic phase as the transition to $N_{\textrm{TB}}$ was approached \cite{Borshch&Kim2013,Adlem&Copic2013}. A Landau--de Gennes mesoscopic theory, where the director is replaced by a symmetric and traceless tensor order parameter field, ${\mathbf{ Q }}$, accounts for a fine structure  of the modulated phases and shows limitations of the director's description \cite{Longa&Trebin1990,Longa&Pajak2016}.

The $N_{\textrm{TB}}$ phase observed has a number of remarkable features. It looks uniform everywhere in space like cholesterics, with a temperature--dependent conical angle, $\theta_\nn$, ranging approximately between 9$^\circ$ and 30$^\circ$ \cite{Borshch&Kim2013,Meyer&Luckhurst&Dozov2015,Sreenilayam&LC&2016}. But, while the cholesteric phase with its conical angle equal to a right angle ($\theta_\nn=\pi/2$), can homogeneously fill the space with twist the analogous arrangement for $N_{\textrm{TB}}$ ($0<\theta_\nn<\pi/2$) requires both bend and twist deformations to be present. X-ray diffraction experiments, sensitive to positional \cite{Cestari&DiezBerart2011,Chen&Porada2013} or orientational \cite{Zhu&Chenhui&2016,Stevenson&2017} orderings reveal no long--range positional order of molecular centers of mass ($N_{\textrm{TB}}$ indeed remains a fully 3D liquid), but a 1D periodic order of molecular orientations. The helicoidal pitch length in the $N_{\textrm{TB}}$ is about 10 nm, {\emph{i.e.}} on the order of a few molecular lengths, which is about two orders of magnitude smaller than that typically found in cholesteric and blue phases \cite{deGennesBook}. The $N_{\textrm{TB}}$ phase is usually stabilized as a result of a first--order phase transition from the uniaxial nematic phase, but very recently a direct transition between $N_{\textrm{TB}}$ and the isotropic phase has also been found \cite{Archbold&Davis2015,Dawood&Grossel2016}. Lack of molecular intrinsic chirality implies that coexisting domains of opposite handedness are formed and, consequently, the emergence of $N_{\textrm{TB}}$ is related to a fundamentally new phenomenon, namely, \emph{the spontaneous chiral symmetry breaking}.

While phenomenologically the spontaneous distortion of the $N_{\textrm{TB}}$ and $N_{\textrm{SB}}$ phases can effectively be explained as originating from the negative bend elasticity \cite{Dozov2001} the question of what microscopic/mesoscopic mechanism can be responsible for chiral symmetry breaking, especially the selforganization into $N_{\textrm{TB}}$, is still open and remains to be understood and explored. The issue has been addressed at the theoretical level in a series of papers \cite{Memmer2002,Shamid&Dhakal&Selinger2013,Greco&Luckhurst&Ferrarini2014,Greco&Ferrarini2015,Osipov&Pajak2016,Vanakaras&Photinos2016,Ferrarini2016,Longa&Pajak2016,Osipov&Pajak2016b,Tomczyk&Pajak2016}. Analysis shows that the molecules whose structure is sufficiently bent is a necessary requirement for the stabilization of $N_{\textrm{TB}}$, probably as a results of entropic, excluded volume interactions \cite{Memmer2002,Greco&Ferrarini2015}. The molecules not only selforganize into a helical structure, but also propagate long--range polar order of vanishing global polarization, transverse to the helical axis. For steric interactions the polarity is a consequence of ordering of molecular bent cores \cite{Greco&Luckhurst&Ferrarini2014,Greco&Ferrarini2015,Osipov&Pajak2016b,Tomczyk&Pajak2016} and the other molecular interactions, like between electrostatic dipoles, are probably less relevant for thermal stability of this phase. These conclusions seem in line with recent experimental observations \cite{Mandle&Davis2014b,Ivsic&2014,Mandle2015}.

A mesoscopic--level explanation of how molecular polarity of bent--core molecules can generate modulated polar phases and, hence, effectively lower the bend elastic constant has been proposed to be due to the flexoelectric couplings, where derivatives of the the alignment tensor (or of the director field), induce a net polarization \cite{Longa&Trebin1990,Shamid&Dhakal&Selinger2013,Longa&Pajak2016,Osipov&Pajak2016,supplemental}. The minimal coupling model, which is able to account for $N_{\textrm{TB}}$, is the Landau--de Gennes (LdeG) type of free energy expansion in the alignment tensor $\Q(\rr)$ and the polarization field $\PP(\rr)$, and their derivatives \cite{Longa&Monselesan&Trebin1987,Longa&Trebin1990,Longa&Pajak2016}. It can be decomposed as
\begin{equation}\label{freeEnergy}
  F\equiv F[\Q,\PP]=\frac{1}{V}\int_V\left( f_{Q}+f_{P}+f_{QP}\right) \mathrm{d}^{3}{\rr},
\end{equation}
where the free energy densities, $f_{X}$, are constructed out of the fields $\{X\}$. By taking suitable units and assuming deformations to appear only in a quadratic part of the free energy, the general form of $f_{X}$ up to fourth order in $X$ for nonchiral liquid crystals is \cite{Longa&Trebin1990}
\begin{eqnarray}  \label{ldegQ}
 \hspace{-0.3cm}
 f_{Q} &=&
    \frac{1}{4} \left[ t_Q\, \mathrm{Tr}({\Q}^{2}) + (\mathbf{\nabla}\otimes \Q)\cdot
       (\mathbf{\nabla}\otimes \Q)  \right.\nonumber\\
     &+&  \left. \rho (\mathbf{\nabla}\cdot \Q)\cdot (\mathbf{\nabla}\cdot \Q) \right]
    - \sqrt{6}\, B\, \mathrm{Tr}({\Q}^{3}) + \mathrm{Tr}({\Q}^{2})^{2},
  \end{eqnarray}
 \begin{eqnarray} \label{ldegP}
  \hspace{-0.3cm}
 f_{P}  &=&
    \frac{1}{4}
       \left[ t_P \, \mathbf{P}^{2} +
              (\mathbf{\nabla \otimes P})\cdot(\mathbf{\nabla \otimes P}) % \right. \nonumber \\
 %      &+& \left.
       + a_c (\mathbf{\nabla \cdot P})^2 \right]\nonumber \\ &+& (\mathbf{P}^{2})^2,
    \end{eqnarray}
  \begin{eqnarray} \label{ldegQP}
   \hspace{-2cm}
 f_{QP} &=&   -\frac{1}{4} e_P \mathbf{\mathbf{P}\cdot( \nabla \cdot Q}) %, \hspace{0.6cm} \\
 %f_{3QP} &=&
 - \lambda P_\alpha Q_{\alpha\beta} P_\beta ,
\end{eqnarray}
where $t_Q$ and $t_P>0$ are the reduced temperatures associated with $\Q$ and $\PP$ fields, respectively; $\rho$ is the relative elastic constant; $a_c$ is the strength of longitudinal contribution from the steric polarization \cite{Longa&Pajak2016}; $e_P$ is the strength of flexopolarization. For thermodynamic stability of this free energy expansion it is also mandatory that $\rho > -\frac{3}{2}$ and $1+a_c > 0$.
%Additionally, $a_4$ must be positive  if $\lambda \ne 0$.
When the flexopolarization coupling becomes strong enough ($\rho-\frac{e_P^2}{4 t_P} \le -\frac{3}{2}$) \cite{Longa&Pajak2016}, the uniform nematic phase can no longer be stable and a modulated polar phase is formed. In addition to $N_{\textrm{TB}}$ and $N_{\textrm{SB}}$ the theory (\ref{freeEnergy}) predicts the existence of further two 1D modulated nonchiral, polar nematic phases with transverse and longitudinal polarization being modulated along just one direction for one star of wave vector approximation \cite{Longa&Pajak2016}. But it is important to observe that the flexopolarization term alone ($e_P \ne 0$) is not sufficient to bring about spontaneous chiral symmetry breaking. It needs to be accompanied, at least, by the nonvanishing $\lambda$--coupling ($\lambda \ne 0$) in $F$. The nonvanishing  $\lambda$ is also needed for a proper explanation of the fluctuation modes in $N_{\textrm{TB}}$, as suggested by recent light scattering experiments \cite{Parsouzi&Shamid2016}.

Alternative mesoscopic scenarios pertaining to the stability of $N_{\textrm{TB}}$, like these involving couplings between the alignment tensor and higher rank (octupolar)  order parameters have also been proposed \cite{Lubensky&Radzihovsky2002,Brand&Pleiner&Cladis2005,Longa&Pajak&Wydro2009,Longa&Pajak2011,Trojanowski&Pajak2012,Longa&Trojanowski2013,Trojanowski&Ciesla&Longa2017}. This indicates that the theoretical studies of mechanism(s) responsible for observed spontaneous chiral symmetry breaking are still in their initial stage and further research is needed to provide understanding of stability of $N_{\textrm{TB}}$. One promising direction, which we would like to follow here, is a systematic analysis of how properties of $N_{\textrm{TB}}$ can change in the presence of external stimuli, such as electric or magnetic fields, surface anchoring, photo--chemically driven trans--cis isomerization \emph{etc.} Such analysis can also be important in seeking for future practical applications of this new phase.

In case of modulated nematics their response to an external field can become highly nontrivial \cite{deGennesBook,Deuling2007}. In cholesterics, for example, it is possible to unwind the orientational spiral through an intermediate heliconical structure \cite{Xiang&Li2015FIELD,Salili&Xiang2016FIELD}, both for bulk sample \cite{Meyer1968} and in confined geometry \cite{Scarfone&Lelidis&Barbero2011,Zakhlevnykh&Shavkunov2016}. A more comprehensible, field--induced modification of cholesterics involves reorientation of helical axis \cite{deGennesBook}, or changing the pitch \cite{WuYang2001}. Similar effects can be expected for $N_{\textrm{TB}}$, although recent magnetic field experiments \cite{Challa&Borshch2014FIELD,Salili&Tamba2016FIELD} show only depletion of the $N-N_{TB}$ transition temperature, without a noticeable distortion of the structure. So far, theoretical attempts to characterize the interaction of $N_{\textrm{TB}}$ with field have been made on the basis of the Frank elastic theory \cite{Meyer2016FIELD,Zola&Barbero2016FIELD,Shiyanovskii&Simonario&Virga2016FIELD}.

A purpose of this paper is to study, in a systematic way, a response of the bulk $N_{\textrm{TB}}$ phase to the external fields (electric, magnetic) within the frame of the LdeG free energy, Eqs.~(\ref{freeEnergy}-\ref{ldegQP}). As $N_{\textrm{TB}}$ is expected for nonchiral bent--shaped molecules, with and without electric dipoles, we assume that stability of this phase is driven primarily by excluded--volume, entropic interactions \cite{Greco&Ferrarini2015}. We consider the LdeG free energy, Eq.~\ref{freeEnergy}, supplemented by the dielectric (diamagnetic) term, $F_E$, ($ F\rightarrow  F+F_{E}$), where
\begin{equation}\label{freeExt}
  F_{E} = - \Delta \epsilon \frac{1}{V}\int_V E_\alpha Q_{\alpha\beta}(\rr) E_\beta\mathrm{d}^{3}{\rr} ,
\end{equation}
and where $\Delta \epsilon$ is the is the dielectric (diamagnetic) anisotropy in the director reference frame. We think that the dielectric (diamagnetic) term should dominate, at least for sufficiently strong fields and disregard a possible direct interaction between the dipole moments and the field. We explore the absolute stability of the $N_{\textrm{TB}}$ phase for the model (\ref{freeEnergy}-\ref{freeExt}) by limiting to a family of all nematic structures, periodic at most in one spatial direction (hereafter referred to as ODMNS \cite{Longa&Pajak2016}). Starting with the $N_{\textrm{TB}}$ phase, which is stable within the ODMNS family for vanishing field, we identify free energy minimizers for nonzero field \cite{Southampton-2016}. A full account of the electric field response, with terms quadratic and linear in field, will be presented elsewhere.

All possible ODMNS structures can be parameterized with the aid of plane waves expansion of $\Q(\rr )$ and $\PP(\rr )$ \cite{Longa&Trebin1990,Longa&Pajak2016}:
%
%\begin{equation} \label{qq}
 $       \Q(\rr) =
         \sum_{n}
               \sum^{2}_{m=-2}
                  {Q_{m}(n) }
                  \exp(\mathrm{i}\, {n k \,\z \cdot\rr} )\,
                  \e^{[2]}_{m,{\z }}
%\end{equation}
$,
%
%\begin{equation}\label{pp}
 $                \PP(\rr) =
         \sum_{n}
               \sum^{1}_{m=-1}
                  {P_{m}(n) }
                  \exp(\mathrm{i}\, {n k \,\z\cdot\rr}
                       %- \mathrm{i}\: \psi_{m,{{\kk}}}
                       )\,
                  \e^{[1]}_{m,{\z }},
$%\end{equation}
$\,$ where ${n k \z = \kk}$  are the wave-vectors ($n=0,\pm 1,...$); $Q_{m}({n})$ and $P_{m}({n})$ are the unknown amplitudes found from the minimization of the free energy expansion, and $\e^{[L]}_{m,{\z }}$, $m=0,\pm1,\pm L$ are the spin $L = 1, 2$ spherical tensors represented in a laboratory coordinate system with quantization axis along ${\z}$-direction. The selection of $k$, $Q_{m}(n)$, and $P_{m}(n)$ is fixed by minimization of $F$, supplemented by the bifurcation analysis \cite{Longa1986,Longa&Grzybowski2005,supplemental}. Note that only $n=0$ terms couple to a uniform external field in (\ref{freeExt}), giving
%
%\begin{widetext}
%\begin{eqnarray}\label{FfieldExplicite}
%F_{E} &=& - \Delta \epsilon E^2 \left\{
%  \sin ^2(\theta ) \left[  x_{20} \cos (2 \phi ) -  y_{20} \sin (2 \phi )\right]
%   +\sin (2 \theta ) \left[  y_{10} \sin (\phi)- x_{10} \cos (\phi )\right] + x_{00} \left[3 \cos (2 \theta )+1\right]/{(2 \sqrt{6})} \right\},
%\end{eqnarray}
%\end{widetext}
%
\begin{eqnarray}\label{FfieldExplicite}
  F_{E} = - \Delta \epsilon E^2 \left\{
  \sin ^2(\theta ) \left[  x_{20} \cos (2 \phi ) -  y_{20} \sin (2 \phi )\right] + \sin (2 \theta )  \right. \nonumber\\
  \left. \left[  y_{10} \sin (\phi) - x_{10} \cos (\phi )\right] +  x_{00} \left[3 \cos (2 \theta )+1\right]/{(2 \sqrt{6})} \right\},
\end{eqnarray}
where $\mathbf{E}=E[ \cos (\phi ) \sin (\theta ), \sin (\phi ) \sin (\theta ) ,\cos (\theta )]$ is the external field expressed in the laboratory system of frame and where $\mathfrak{Re}Q_m(n)=x_{mn}$ and $\mathfrak{Im}Q_m(n)=y_{mn}$. The relative orientation of $\Q$ and $\mathbf{E}$, parameterized by $\theta$ and $\phi $, is found by minimization of $F$. The formulas (\ref{ldegQ}-\ref{ldegQP}) entering $F$ are given in the expanded form in \cite{supplemental}.

Our starting point is the identification of  homogeneous structures of wave vector $k$ that can be constructed out of $\Q$ and $\PP$, among which should be the $N_{\textrm{TB}}$ phase. The spatial homogeneity of a structure implies that $\forall z$ the tensors $\Q(z)$ and, say, $\Q(0)$ for $z=0$ are connected by a rotation. More specifically, $\Q(z)$ can be obtained from $\Q(0)$ by rotating through the angle $kz$ about $\mathbf{\hat{z}}$: $\mathcal{R}_{\hat{z}}(\pm k z) \Q(0) = Q_{\pm }(z)$, where $\pm$ stands for right-- and left--handed heliconical structure. Likewise, through the same rotation $\PP(0)$ is transformed into $ \PP(z)$: $ \mathcal{R}_{\hat{z}}(\pm k z) \PP(0) = \PP_{\pm}(z)$. The structures fulfilling the above conditions have a \emph{unique} form given by the terms of $|n| \le 2$ in the expansions of $\Q$ and $\PP$, namely
\begin{eqnarray} \label{RQ1}
 &\Q_{\pm}(z) = Q_0(0) \e^{[2]}_{0,{\z }} + \sum_{m=1}^{2} \left[Q_{\pm m}(m) \e^{[2]}_{\pm m,{\z }}  e^{\pm i m k z}  + c.c \right]\!\!,\;\; \\\label{RP1}
 &\PP_\pm(z) = P_0(0) \e^{[1]}_{0,{\z }} + \left[P_{\pm 1}(1)  \e^{[1]}_{\pm 1,{\z }}  e^{\pm i k z}  + c.c \right]\!\!.
\end{eqnarray}
For example, the explicit formulas for $\Q_+$ and $\PP_+$ read
%
%\begin{strip}
\begin{widetext}
\begin{eqnarray}\label{QNTBrepresentation}
  \hspace{-5mm}\Q(z)\equiv \Q_+(z)&=&
\left[
\begin{array}{ccc}
 \cos (2 k z+\phi _2) r_2-\frac{x_{00}}{\sqrt{6}} & -\sin (2 k z+\phi
   _2) r_2 & -\cos ( k z+\phi _1) r_1 \\
 -\sin (2 k z+\phi _2) r_2 & -\cos (2 k z+\phi _2)
   r_2-\frac{x_{00}}{\sqrt{6}} & \sin (k z+\phi _1) r_1 \\
 -\cos (k z+\phi _1) r_1 & \sin (k z+\phi _1) r_1 &
   \sqrt{\frac{2}{3}} x_{00} \\
\end{array}
\right],\;
\end{eqnarray}
\end{widetext}
%\end{strip}
%
\begin{eqnarray}\label{PNTBrepresentation}
  \PP(z)\equiv \PP_+(z)&=& \left[
 \begin{array}{c}
 -\sqrt{2} p_1 \cos \left(k z +\phi _p\right)\\
 \sqrt{2} p_1 \sin \left(k z + \phi
   _p\right)\\
   v_{00}
   \end{array}
   \right].
\end{eqnarray}
Here $r_i \cos \left(\phi _i\right)=x_{ii}$, $r_i \sin \left(\phi _i\right)=y_{ii}$ ($r_i \ge 0$), $ p_1 \cos \left(\phi _p\right)=v_{11}$, $p_1 \sin \left(\phi_p\right)=z_{11}$ ($p_1 \ge 0$), where $v_{ij}=\mathfrak{Re}P_i(j)$, $z_{ij}=\mathfrak{Im}P_i(j)$  and, as previously, $x_{ij}=\mathfrak{Re}Q_i(j)$ and $y_{ij}=\mathfrak{Im}Q_i(j)$. Note that one of the three phases $\phi_i$, ($i=1,2,p$) is the Goldstone mode and can be eliminated, which expresses the freedom of choosing the origin of the laboratory system of frame. \newpage \noindent In what follow we set $\phi_1=0$. Hence, a family of all homogeneous (polar) helical nematic mesophases, structurally linked with the uniaxial nematic phase ($N$), can be parameterized unambiguously using at most eight parameters (including $k$ vector). The $x_{00}$ terms in Eq.~(\ref{QNTBrepresentation}) represents the reference $N$ phase with the director along ${\z }$. As the systems we study are nonchiral, the spontaneous chiral symmetry breaking means that domains representing opposite chiralities have the same free energy: $F[\Q_+,\PP_+]=F[\Q_-,\PP_-]$ and that they are formed with the same probability (ambidextrous chirality).

Setting $r_{1}=0$ in Eq.~(\ref{QNTBrepresentation}) gives the cholesteric phase of the conical angle $\theta_\nn=\pi$, Eq.~(\ref{nn}), while the simplest of the $N_{\textrm{TB}}$ phases is obtained by neglecting terms with $m=2$  ($r_2=0$) in (\ref{QNTBrepresentation}). In this simplified case the conical angle is given by
\begin{equation}\label{NTBT0}
  \cos(\theta_\nn)=\frac{\sqrt{3 \chi ^2+8}+\sqrt{3} \chi }{\sqrt{2} \sqrt{3 \chi ^2+\sqrt{9 \chi
   ^2+24} \chi +8}},
\end{equation}
where $\chi=x_{00}/r_1$. Note that $0 \le \theta_\nn \le \pi/4$ for prolate uniaxial nematic background ($x_{00} \ge 0$), while the oblate case ($x_{00} \le 0$) yields $\pi/2\ge \theta_\nn \ge \pi/4$. The general case of $N_{TB}$ with $r_2\ne0$ allows for a fine control of biaxiality and of the conical angle. For example, the biaxiality of $N_{\textrm{TB}}$, measured by the normalized parameter $w$ \cite{Allender&Longa&2008}
\begin{equation}\label{biaxialityPar}
  -1\le w(\Q)={\sqrt{6}\: Tr(\Q^3)}/{\left[ Tr(\Q^2)  \right]^{\frac{3}{2}}} \le 1\; ,
\end{equation}
is given by
\begin{equation}\label{biaxNTB}
w=  \frac{\chi  \left(-6\tau ^2+\chi ^2+3\right)+3 \sqrt{6}\:\tau  \cos \left(\phi
   _2\right)}{\left(2\tau ^2+\chi ^2+2\right)^{3/2}}.
\end{equation}
Here $\tau =r_2/r_1$ and $w=1$ ($w=-1$) for local uniaxial prolate (oblate) order. The states of $|w|<1$ are biaxial with $w=0$ corresponding to maximally biaxial order. Note that $w$, Eq.~(\ref{biaxNTB}), is $k$--independent, expressing the fact that $z$--dependence of $\Q$, Eq.~(\ref{QNTBrepresentation}), is generated by a rotation. This means that $N_{\textrm{TB}}$ is biaxial and in the limit of $k \to 0$ we get the fourth of homogeneous nematic structures accounted for by (\ref{QNTBrepresentation}), namely the biaxial nematic.

Each of the structures identified so far can be polar, Eq.~(\ref{PNTBrepresentation}). For $k \ne 0$ the polarization can acquire the long--range periodic component in the $x-y$ plane ($p_1\ne0$), which is perpendicular to $\z$ and/or global macroscopic polarization ($v_{00}\ne 0$), parallel to  ${\z }$ . Interestingly, the $N_{\textrm{TB}}$ phase given by Eqs~(\ref{QNTBrepresentation},\ref{PNTBrepresentation}) with $v_{00}= 0$, can be the global minimizer within the ODMNS class. The sufficient condition for the free energy parameters can be derived using the bifurcation analysis \cite{supplemental} with uniaxial nematic $(x_{00} \neq 0)$ as a reference state. It reads
\begin{equation}\label{bifToNTB}
  \frac{-2 \sqrt{6} t_P}{\sqrt{9 B^2-8 t_Q}+3 B}<\lambda <\frac{\sqrt{\frac{3}{2}} \left(e_P^2-4 (\rho +2) t_P\right)}{(\rho +2) \left(\sqrt{9 B^2-8 t_Q}+3 B\right)},
\end{equation}
which is valid when the following additional constrains are fulfilled: $e_P \neq 0$, $t_Q<B^2$, $\rho>-\frac{3}{2}$ and $t_P>0$.

In seeking for a globally stable structure among ODMNS we need to take into account both homogeneous and inhomogeneous trial states given by the plane waves expansion of $\Q$ and $\PP$. The complexity of the ODMNS minimization depends on the number of amplitudes used in this expansion, which is controlled by the maximal value $n_{max}$ of $|n|$ in the set $\{k, Q_m(n), P_m(n),\; n=0,\pm 1, ...,\pm n_{max} \}$. As it turns out the $n_{max}=1$ approximation with 25 \emph{real}, variational parameters is not sufficient to qualitatively reproduce structural changes as experienced by $N_{\textrm{TB}}$ due to an applied external field. In order to obtain credible results the following strategy has proved to work. In the first step we perform the free energy minimization with $n_{max}=2$. Then, the identified structures serve in the next step as initial states in seeking for the improved free energy minimum, where relaxation method for 1D periodic structures is being used.

More specifically, we consider the bulk $N_{\textrm{TB}}$ sample of thickness $\Lambda$ with free boundary conditions at $z=0$ and $z=\Lambda$ ($0\le z \le \Lambda $), where $\Lambda$ is much larger than the period of the structure. Then, we divide the sample in the $z$--direction into $N$ equal intervals and approximate the fields $\Q(z)$ and $\PP(z)$ at nodes $z_i$ taken in the middle of each interval $i$. For the derivatives of the fields we use the central difference approximation
\begin{equation}
\left. \frac{\partial \mathbf{X}(z)}{\partial z} \right|_{ z = z_i }=\frac{\mathbf{X}(z_{i+1})-\mathbf{X}(z_{i-1})}{2\delta},
\end{equation}
where $\delta$ is the distance between the two neighboring nodes and where $\mathbf{X}= \Q(z)$ or $\PP(z)$. After these preparations are done the volume integral in $F$ is discretized in a standard way using simple trapezoidal rule, with the node variables  $\{\mathbf{X}\}\equiv\{\Q(z_i)$,$\PP(z_i)$, $i=1,...8N\}$ playing the role of variational parameters.

The relaxation method for the discretized free energy determines $\{\mathbf{X}\}$ with the following iterative formula
\begin{eqnarray}
\mathbf{X}^{(n+1)}(z_i)-\mathbf{X}^{(n)}(z_i)
=-\gamma
\left.\frac{\partial F}
{\partial \mathbf{X}(z_i)}\right|_{ \mathbf{X}(z_i)= \mathbf{X}^{(n)}(z_i)},
\label{relax}
\end{eqnarray}
where the superscript $(n)$ enumerates the values of $\{\mathbf{X}(z_i)\}$ obtained in successive iterations and $\gamma$ is the relaxation parameter. For the convergent iterations  we typically used $\gamma=0.005$ and  $N=4000$ but, generally, the choice of $\gamma$ is not important, unless numerical instabilities occur. Occasionally, we doubled the number of nodes to see the influence on the accuracy of the free energy. The iteration was initialized with $\mathbf{X}^{(1)}(z_i) \equiv \mathbf{X}(z_i)$ using analytic expressions obtained from the Fourier amplitudes minimization for $n_{max}=2$ for the case without the external field. The corresponding periodicity ($2 \pi/k$) of $N_{\textrm{TB}}$ was used to fix $\Lambda$. Once initialized, the system (\ref{relax}) was solved iteratively until selfconsistency with the required accuracy was achieved. We used very large bulk samples of $\Lambda \approx 260 \pi/k$ for which the structure close to the midplane ($z \approx \Lambda/2$) was insensitive to the ordering near boundaries. The next step was the equilibration of the $k$ vector for the $\mathbf{X}$ fields evolved during the relaxation process. The method we used is described in detail in \cite{Alexander&Yeomans2006,HenrichBPS} and is based on the observation that under distance rescaling: $z\to  z/\kappa$ the free energy is a general quadratic function of $\kappa$, with coefficients that are $\kappa$-independent. Thus the scaling factor, $\kappa^*$, obtained by the subsequent minimization of the free energy, gives the improved wave vector, $\kappa^* k$, for the approximate $\mathbf{X}$ fields that are obtained from relaxation. Effectively, it amounts in rescaling the distance between the nodes: $\delta\to\delta'=\delta/\kappa^*$ of the relaxed fields. The relaxation procedure is then repeated with the new $\delta'$ and followed by a new rescaling. The process continued until $ \kappa^* \cong 1$. To find out solutions for the cases with electric field we repeated the above procedure using as the initial condition the solution from the step with a smaller field.

\begin{figure}[h!]
\centering
\includegraphics[scale=0.61]{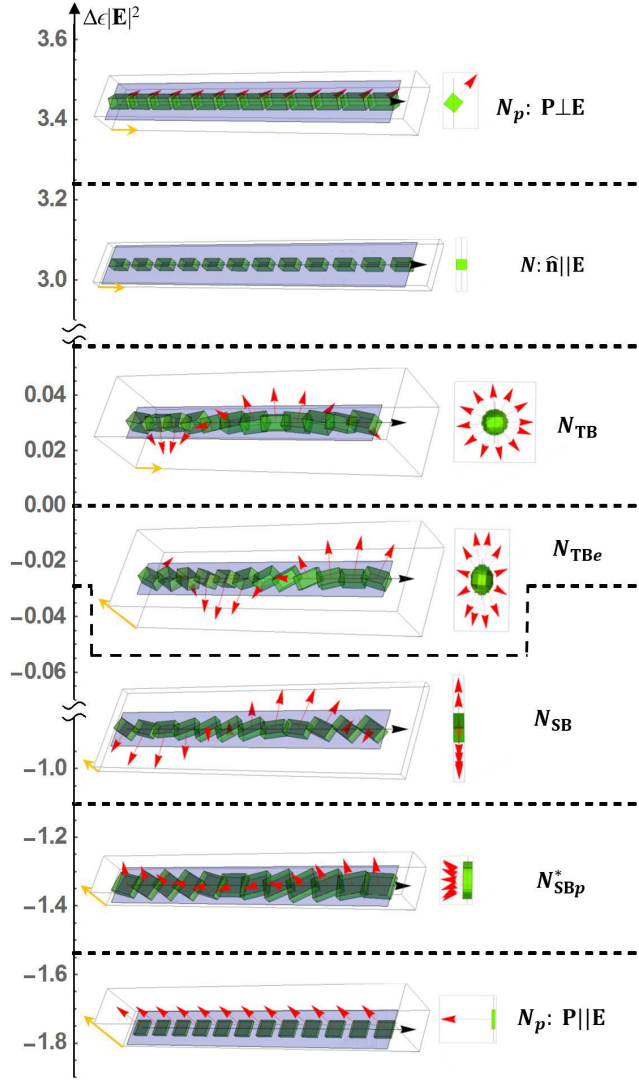}
\caption{(Color online) Sketch of $N_{\textrm{TB}}$ modifications under the external field. Dashed lines correspond to the strength of external field coupling that leads to a phase transformation; yellow arrows are directions of $\mathbf{E}$ in each of the structures. Red arrows represent $\PP$ and  black arrow is the direction of $\kk$. Lengths of cuboid edges are proportional to the eigenvalues of $\Q + c \mathbf{I}$, where $\mathbf{I}$ is the unit matrix and $c$ is a constant, such that the isotropic state is represented by a cube. Every phase is presented in a perspective with its top view added. }
\label{Fig-structures}
\end{figure}

Detailed analysis is performed for the case when $\rho=1$, $a_c=2$, $e_P=-4$, $B=1/\sqrt6$, $\lambda=-1/2$, $t_Q=1/10$ and $t_P=8/10$ where without external field the heliconical structure gives minimum among ODMNS, isotropic, uniaxial and biaxial nematics. Depending on the strength of the field and the sign of the material anisotropy, both controlled by the single model parameter $\Delta\epsilon E^2$ in Eq.~(\ref{freeExt}), the $N_{\textrm{TB}}$ structure evolves as shown in Fig.~(\ref{Fig-structures}).
For the case of positive material anisotropy ($\Delta\epsilon E^2>0$) the minimum is realized for $\mathbf{E} || \hat{\kk}$, while for negative anisotropy ($\Delta\epsilon E^2<0$) the minimum occurs when $\mathbf{E} \bot \hat{\kk}$. Furthermore, in the positive anisotropy case the $N_{\textrm{TB}}$ phase unwinds to the uniaxial nematic with the director parallel to the external field. Further increase of the field results in stabilization of a polar nematic ($N_p$), where secondary director is parallel to the polarization vector, both being perpendicular to the external field. The case of negative anisotropy modifies instantly the modulation of the main director in the $N_{\textrm{TB}}$ phase, which is now precessing on the right elliptic cone around $\hat{\kk}$. We denote this new structure as $N_{\textrm{TB} e}$. Stronger fields ($\mathbf{E} \bot \hat{\kk}$) make the elliptic cone base narrower and finally $N_{\textrm{SB}}$ is stabilized, where
 the cone becomes degenerated to a line and $\PP$ lies in the plane of splay--bend modulations. With larger fields in--plane modulations diminish and $\PP$ acquires the off-plane uniform component along the field. This is a new chiral structure denoted $N^*_{\textrm{SB} p}$. For even stronger fields this phase transforms to polar nematic with polarization parallel to the field.

Quantitatively phases are described by the amplitudes $Q_{m}({n})$ and $P_{m}({n})$, or equivalently by their real and imaginary parts, which we denoted: $x_{ij}=\mathfrak{Re}Q_i(j)$, $y_{ij}=\mathfrak{Im}Q_i(j)$, $v_{ij}=\mathfrak{Re}P_i(j)$, $z_{ij}=\mathfrak{Im}P_i(j)$. Additionally each structure is characterized by the wave vector $k$. For stable ODMNS with $n_{max}=2$ the sets of nonzero parameters are: $\{y_{11}, y_{22}, x_{00}, x_{11}, x_{22}, z_{11}, v_{11}\}$ for $N_{\textrm{TB}}$; as in $N_{\textrm{TB}}$ and $\{y_{-22}, y_{-11}, y_{02}, x_{-22}, x_{-11}, x_{02}, x_{20}, z_{-11}, z_{02}, v_{-11}, v_{02}\}$ for $N_{\textrm{TB} e}$; as in $N_{\textrm{TB} e}$, provided that the following constrains are fulfilled $\{ |y_{-22}|=|y_{22}|, |y_{-11}|=|y_{11}|, |x_{-22}|=|x_{22}|, |x_{-11}|=|x_{11}|, |z_{-11}|=|z_{11}|, |v_{-11}|=|v_{11}|\}$ for $N_{\textrm{SB}}$; and as in $N_{\textrm{SB}}$ in union with $\{ |y_{-21}|=|y_{-12}|, |y_{12}|=|y_{21}|, |x_{-21}|=|x_{-12}|, |x_{12}|=|x_{21}|, |z_{-12}|=|z_{12}|, z_{10}, |v_{-12}|=|v_{12}| \}$ for $N^*_{\textrm{SB} p}$.
%The amplitudes can be used to calculate $\Q$ and $\PP$ fields.
%
\begin{figure}[htb]
\vspace{0.65cm}
\includegraphics[scale=0.42]{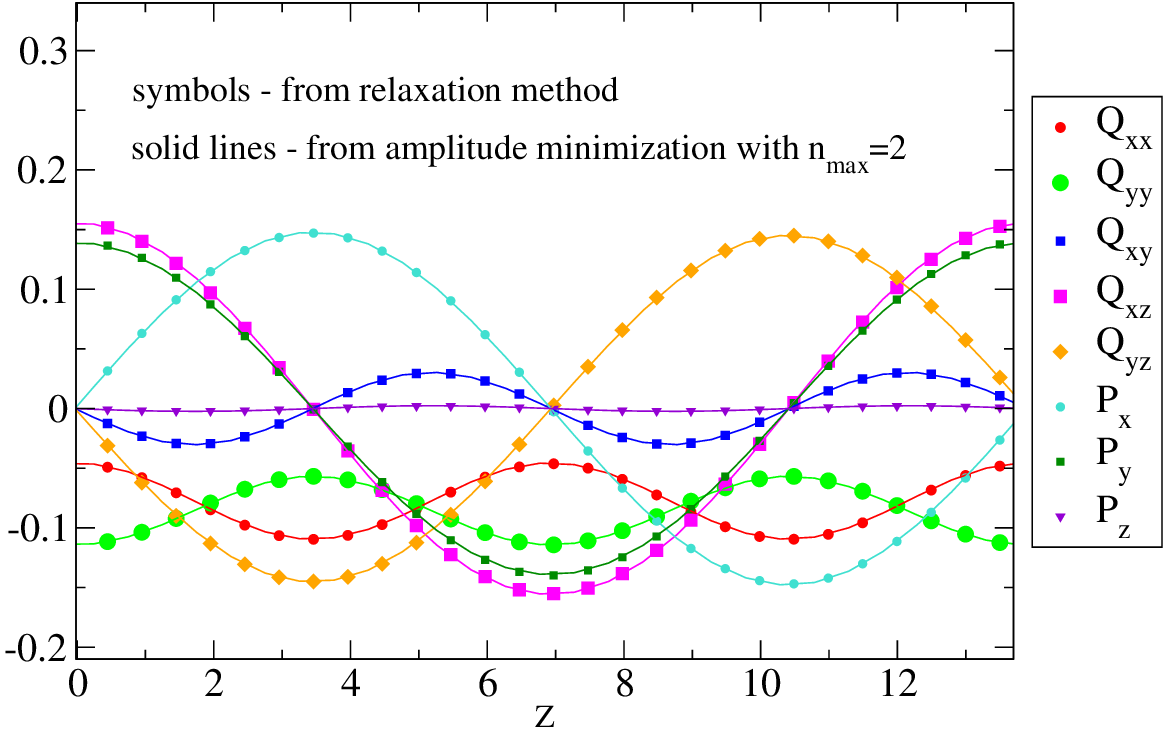}
%\vspace{1cm}
\caption{(Color online) ${\Q}$ and ${\PP}$ fields obtained for $N_{\textrm{TB}}$ from the relaxation method (symbols) and from the free energy minimization with respect to the Fourier amplitudes for $n_{max}=2$ (lines) for $\Delta\epsilon E^2=0.0025$ .}
%\vspace{1cm}
\label{relaxation1}
\end{figure}
\begin{figure}[htb]
\vspace{0.25cm}
\includegraphics[scale=0.42]{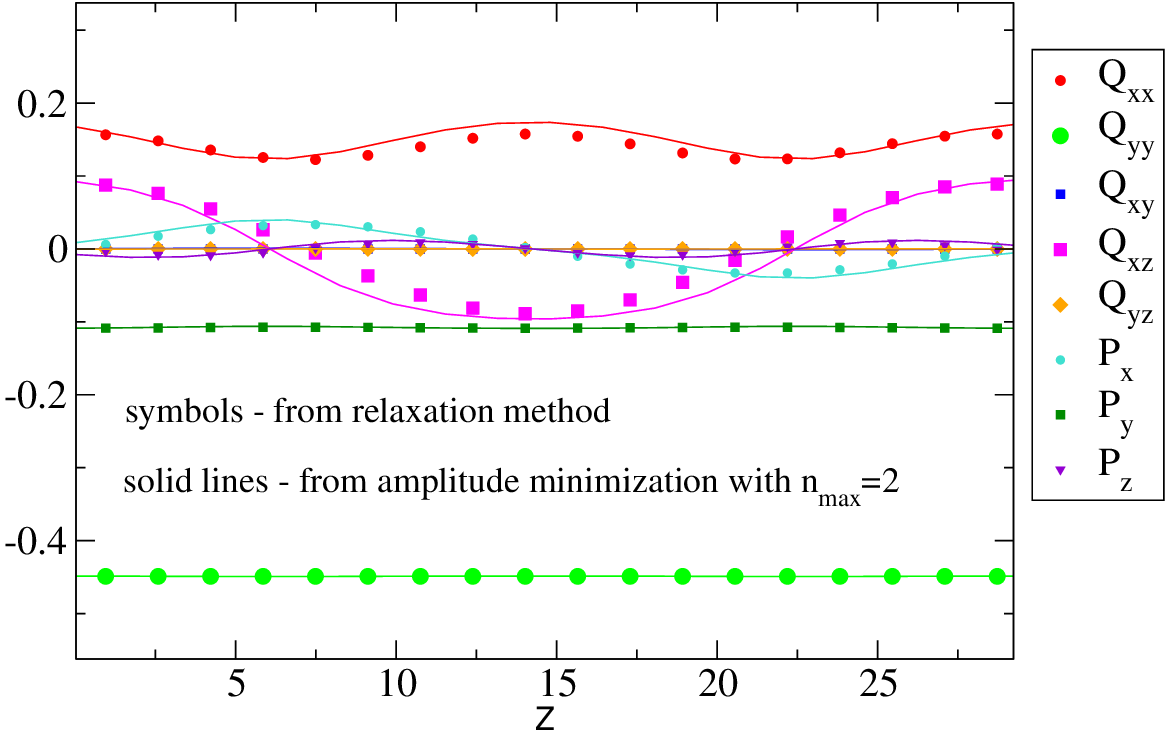}
%\vspace{1cm}
\caption{(Color online) ${\Q}$ and ${\PP}$ fields obtained for $N^*_{\textrm{SB} p}$ from the relaxation method (symbols) and from the free energy minimization with respect to the Fourier amplitudes for $n_{max}=2$ (lines) for $\Delta\epsilon E^2=-1.3225$ . }
\label{relaxation2}
\end{figure}

\begin{figure}[htb]
\hspace{0.75cm}
\includegraphics[scale=0.51]{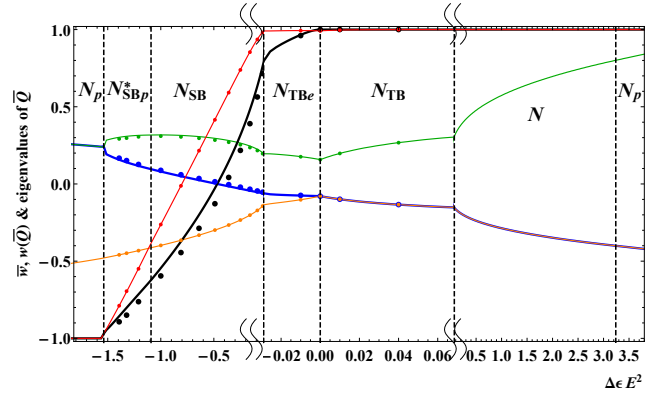}
\vspace{0.25cm}
\caption{(Color online) Various observables averaged over one period from the relaxation method (points) and from the free energy minimization with respect to the Fourier amplitudes for $n_{max}=2$ (lines). Eigenvalues of $\overline{\textrm{\textbf{\emph{Q}}}}$ are given in green (along wave vector), orange (for negative anionotropy it is along the direction of external field) and blue. Biaxiality parameter, given by Eq. (\ref{biaxialityPar}), is plotted in red for $\overline{w}$ and in black  for $w(\overline{\textrm{\textbf{\emph{Q}}}})$. }
\label{Fig-eigenvalues}
\end{figure}

\begin{figure}[htb]
\hspace{0.75cm}
\includegraphics[scale=0.54]{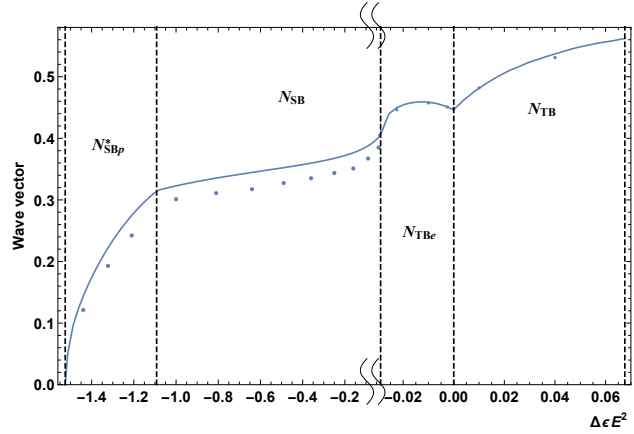}
\vspace{0.25cm}
\caption{(Color online) Wave vector as a function of external field for ODMNS presented in Fig.~(\ref{Fig-structures}). As previously points are from relaxation method and line is outcome of the free energy minimization with respect to the Fourier amplitudes for $n_{max}=2$. }
\label{Fig-wave-vector}
\end{figure}

\begin{figure}[htb]
\hspace{0.85cm}
\includegraphics[scale=0.51]{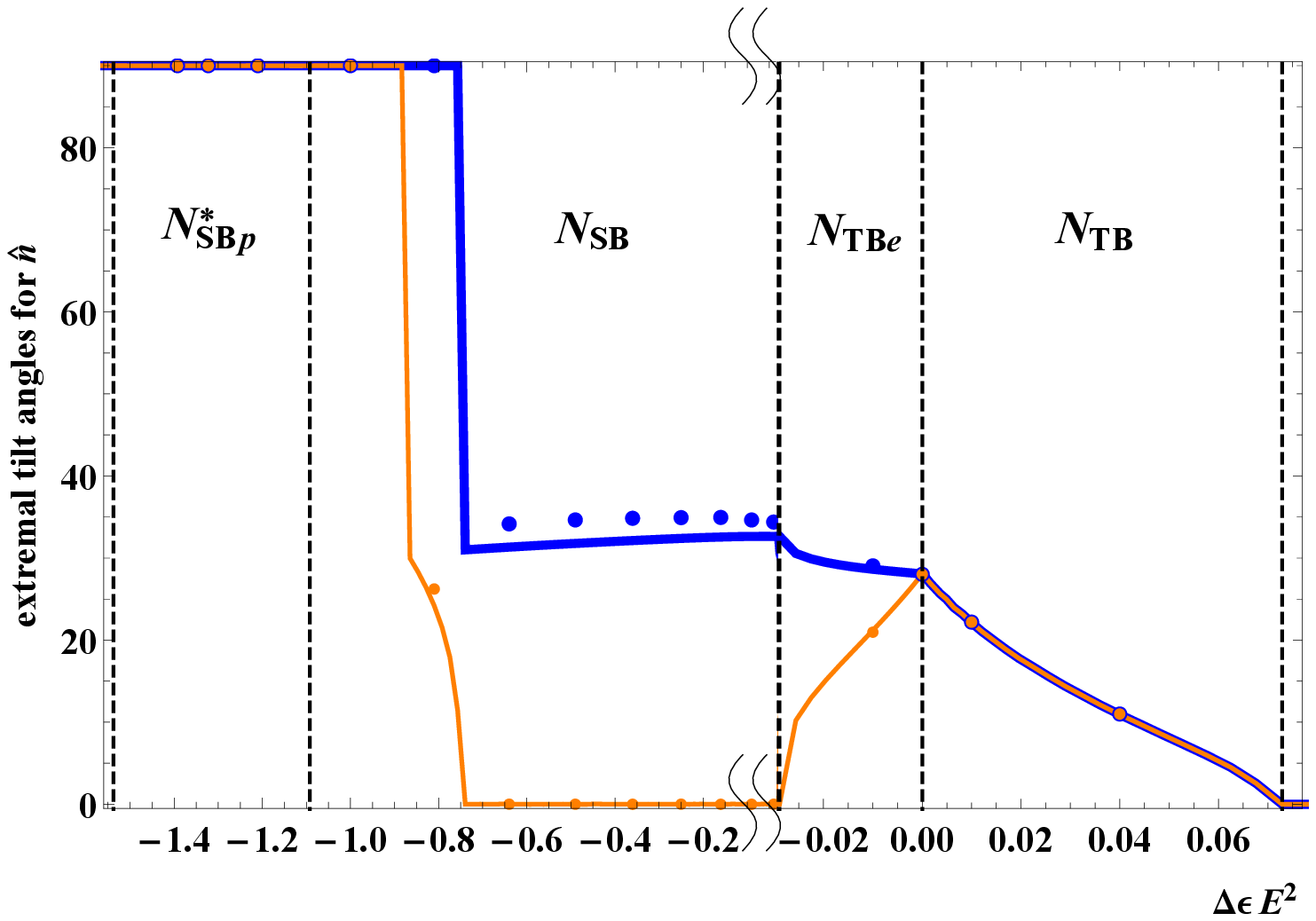}
\vspace{0.25cm}
\caption{(Color online) Minimal (orange) and maximal (blue) values of angle between $\kk$ and $\nn$. As previously points are from relaxation method and lines are outcomes of the free energy minimization with respect to the Fourier amplitudes for $n_{max}=2$. }
\label{Fig7}
\end{figure}

\begin{figure}[htb]
\hspace{0.85cm}
\includegraphics[scale=0.51]{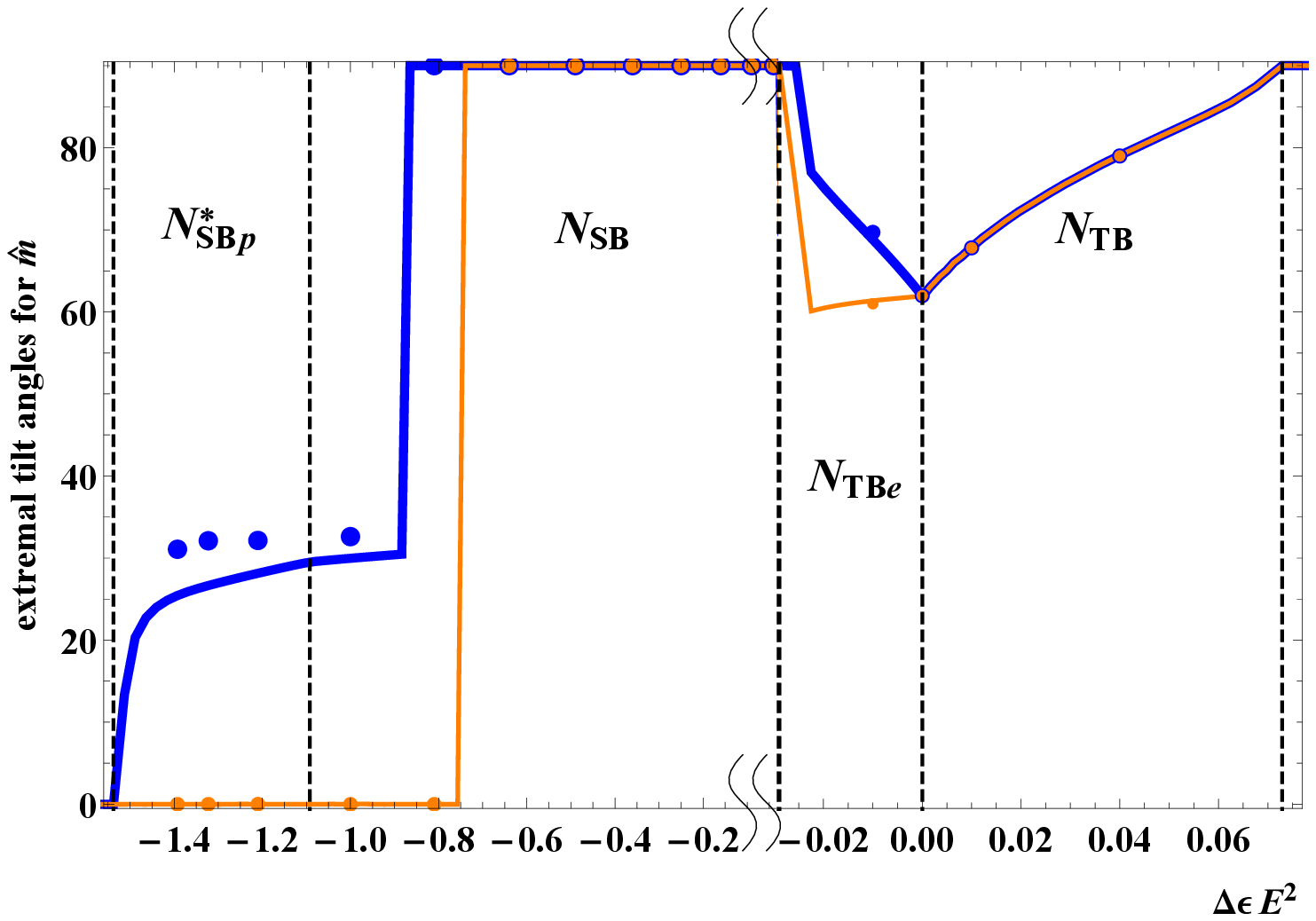}
\vspace{0.25cm}
\caption{(Color online) Minimal (orange) and maximal (blue) values of angle between $\kk$ and $\mm$. As previously points are from relaxation method and lines are outcomes of the free energy minimization with respect to the Fourier amplitudes for $n_{max}=2$. }
\label{Fig8}
\end{figure}

\begin{figure}[htb]
\hspace{0.75cm}
\includegraphics[scale=0.51]{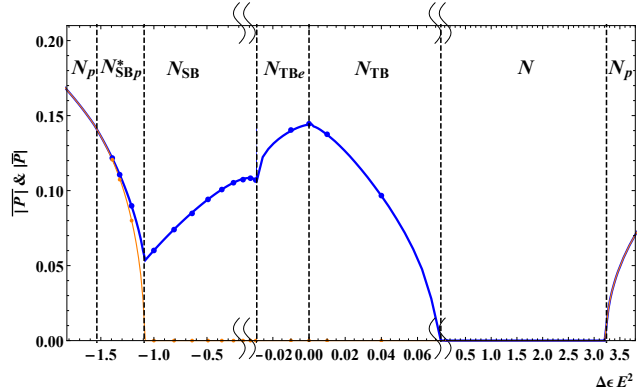}
\vspace{0.25cm}
\caption{(Color online) Averages over one period of $\overline{\textrm{|\textbf{\emph{P}}|}}$ (blue) and $|\overline{\textrm{\textbf{\emph{P}}}}|$ (orange). As previously points are from relaxation method and lines are outcomes free energy minimization with respect to the Fourier amplitudes for $n_{max}=2$. }
\label{Fig9}
\end{figure}

A typical outcome of the relaxation procedure, Eq. (\ref{relax}), for stable structures and how it compares with the amplitude minimization for $n_{max}=2$ is illustrated in Figs~(\ref{relaxation1}-\ref{Fig9}). Note that both the $n_{max}=2$ amplitude minimization and the relaxation results agree quantitatively, indicating that the approximation of $n_{max}=2$ is sufficient to obtain the basic features of the phases studied. Taking $n_{max}=1$ leads to qualitatively inconsistent results.


In order to give an insight into  fine structure of stable phases we plot characteristic observables
for each of them in Figs~(\ref{Fig-eigenvalues}-\ref{Fig9}).
Firstly, we present the behavior of eigenvalues for $\mathbf{Q}(z)$ averaged over one period ($\overline{\textbf{\emph{Q}}}$), Fig. (\ref{Fig-eigenvalues}), which is what can effectively be  measured in experiments. Please note that homogenous nematics and $N_{\textrm{TB}}$ (averaged over one period) are uniaxial, as two eigenvalues of $\overline{\textbf{\emph{Q}}}$ coincide, and all other ODMNS are biaxial. Degree of biaxiality can be quantified by the relative differences between the eigenvalues, or with the help of the $w$ parameter, Eq. (\ref{biaxialityPar}). In Fig. (\ref{Fig-eigenvalues}) we present  $w(\overline{\textbf{\emph{Q}}})$
 in black, and $\overline{w}$ in red, and one sees that $N_{\textrm{TB} }$ ($\overline{w}=$) and  $N_{\textrm{TB} e}$ are weakly biaxial, while $N_{\textrm{SB}}$ is always biaxial passing the point of maximal biaxiality.  Finally $N_{\textrm{SB} p}$  is biaxial of oblate type and $N_p$ is uniaxial oblate.
 Further important characteristic of ODMNS presented in Fig.~(\ref{Fig-structures}) is the variation of the wave vector with field, which is shown in Fig.~(\ref{Fig-wave-vector}). Clearly, the wave vector vanishes in homogenous nematic phases.
 Note that the general effect of the field is to unwind the structures except for the initial behaviour of  $N_{TBe}$, which is just opposite.


Another variable which characterizes   ODMNS  is the conical angle. This tilt angle is measured between $\kk$ and $\nn$, but could
also be given {\em e.g.} between  $\kk$ and $\mm$.  Being constant for $N_{TB}$ it varies with $z$ for all remaining ODMNS,
as shown in Fig.~(\ref{Fig7}), where we depict its minimal and maximal value with field.
 It is apparent that both semiaxes of ellipse are equal in $N_{\textrm{TB}}$ (blue and orange lines in  Fig.~(\ref{Fig7}) coincide), but higher field makes the diameter of cone's basis smaller. In $N_{\textrm{TB} e}$ blue and orange lines split providing the elliptic profile of the cone's basis. In the $N_{\textrm{SB}}$ phase the minimal value of that angle is equal to zero and $\mm$, Fig.~(\ref{Fig8}), is perpendicular to $\kk$,  which means that $\nn$ performs in--plane modulations between minimal and maximal values of $\theta_\nn$. Then, comparing Fig.~(\ref{Fig7}) with Fig.~(\ref{Fig8}) for the angle between $\kk$ and $\mm$, gives insight into changing primary director from $\nn$ to $\mm$ as field increases for materials with negative anisotropy. This effect occurs when $\Delta\epsilon E^2\cong-0.8$ and it is in fact caused by passing through a point of maximal biaxiality from prolate-- to oblate--like structure, as shown in Fig.~(\ref{Fig-eigenvalues}) for $\overline{w}$.
Finally,  in Fig.~(\ref{Fig9}) we present the averages over one period for the length of polarization vector. Average of polarization modulus shows the average length of $\PP$ for each structure, whereas modulus of averaged polarization reveals which of the phases posses the uniform component of $\PP$.
% Only in phases with nonzero constant $\PP$ it is feasible to consider linear coupling between polarization and external field.

Scientists have long sought to understand how chiral states can be generated in a liquid state from nonchiral matter. Now strong evidence is found that a new class of nematics called a nematic twist--bend provides such an example. This entropically induced state is realized because the underlying molecules have a specific shape. Here we present possible transformations of the $N_{\textrm{TB}}$ phase with an external field within the LdeG theory of flexopolarization. The outcomes of numerical minimization along with the full free energy relaxation method have been studied for few sets of model parameters, with one typical
example being  discussed in depth here.

For materials with positive  anisotropy the unwinding of the helix to the uniaxial nematic structure is obtained, however here for sufficiently strong fields a polar nematic appears more stable. Materials with negative anisotropy give rise to  three different ODMNS with a wide range of the $N_{\textrm{SB}}$ phase, so it is an apparent suggestion that this phase can be stabilized  when applying external fields, and the experimental possibility for such an option has been presented before \cite{ECLCFIELD}. So far the $N_{\textrm{SB}}$ was expected as an intermediate state between neighboring domains of opposite chirality \cite{Meyer&Luckhurst&Dozov2015}, which permits for smooth transition between adjacent right-- and left--handed $N_{\textrm{TB}}$ domains.

\acknowledgments{This work was supported by the Grant No. DEC-2013/11/B/ST3/04247 of the National Science Centre in Poland. GP acknowledges also support of Cracowian Consortium `$,\!,$Materia-Energia-Przysz\l{}o\'s\'c'' im. Mariana Smoluchowskiego' within the KNOW grant.}

%\showacknow{} % Display the acknowledgments section

% \pnasbreak splits and balances the columns before the references.
% Uncomment \pnasbreak to view the references in the PNAS-style
% If you see unexpected formatting errors, try commenting out \pnasbreak
% as it can run into problems with floats and footnotes on the final page.
%\pnasbreak

% Bibliography
\bibliography{pnas-sample}

\end{document}